\documentclass[11pt, oneside]{article}   	
\usepackage[margin=1in]{geometry}  
\usepackage{graphicx}				
										
\usepackage{amssymb}
\usepackage{physics}
\usepackage{amsmath}
\usepackage[dvips]{epsfig}

\usepackage{amsfonts}
\usepackage{subfig}

\usepackage{mathrsfs}

\usepackage{xcolor}

\usepackage{bm}

\usepackage{blindtext}

\usepackage[title]{appendix}

\usepackage{cite}

\usepackage{authblk}

\def\cchi{\raise2pt\hbox{$\chi$}} 

\title{\bf{Transient fields in oblique scattering from an infinite planar dielectric interface - a qubit lattice simulation}}

\author {Min Soe ${}^1$, George Vahala ${ }^{2}$, Linda Vahala ${ }^{3}$, Efstratios Koukoutsis ${ }^{4}$, Abhay K. Ram ${ }^{5}$, Kyriakos Hizanidis ${ }^{4}$\\
${ }^{1}$ Department of Mathematics and Physical Sciences, Rogers State University, Claremore,OK 74017\\
${ }^{2}$ Department of Physics, William \& Mary, Williamsburg, VA23185 (retired) \\
${ }^{3}$ Department of Electrical \& Computer Engineering, Old Dominion University, Norfolk, VA 23529\\
${ }^{4}$ School of Electrical and Computer Engineering, National Technical University of Athens,Zographou 15780, Greece \\
${ }^{5}$ Plasma Science and Fusion Center, MIT, Cambridge, MA 02139\\}
\date{}

\begin{document}
\maketitle
\begin{abstract}
An initial value algorithm is utilized to examine the time dependent evolution of the elec-
tromagnetic fields arising from oblique scattering of bounded pulses from an infinite planar
dielectric interface. Since the qubit lattice algorithm (QLA) is almost fully unitary, one finds
excellent conservation of electromagnetic energy. Various Gaussian envelope pulses are consid-
ered in regimes where the incident angle is below that needed for total internal reflection. While
the reflected pulse retains its overall Gaussian shape, the transmitted pulse exhibits a combi-
nation of a Gaussian envelope along with Huygen-like emitted wave fronts from the collision
point of the initial pulse with the infinite dielectric interface. The strength of these Huygen
wavefronts depends on the width of the incident pulse.

\end{abstract}
\section{Introduction}

\qquad The understanding and transient behavior of electromagnetic wave propagation 
and scattering
in inhomoge- neous plasmas is of fundamental importance in magnetic fusion and solar physics as well as other branches of plasma physics. Of particular interest is the study of wave cut-offs, resonance and mode conversion within the plasma from the viewpoint of an initial value problem and without the im- position of internal boundary conditions. One of the simplest plasma models is a two-species warm plasma with its dynamics given by the continuity and momentum equations with simple closure using ideal gas law. Coupling this model with the Maxwell equations permits a simple standard theory of linear wave propagation to be established through the resulting dielectric tensor. In this paper we wish to start on this road using a qubit lattice algorithm (QLA). In principle, QLA [1-6] is a quantum-inspired discrete lattice algorithm built on an interleaved sequence of specially chosen unitary collision and streaming operators that act on the qubit amplitudes, recovering the desired continuum physics to second order in the lattice spacing. It is important to realize that QLA is not a direct finite difference representation of the field equations themselves. As a quantum algorithm, the QLA representation is ideal for direct encoding on a quantum computer [7-13]. Moreover, it
exhibits ideal parallelization to all available cores on classical supercomputers since the unitary collisions act only
on  local data at each lattice site while the streaming operators are simple shift operations. This duality makes QLA a very interesting and important initial value algorithm. At the quantum computation level, the unitary collisions entangle qubit amplitudes while the unitary streaming spreads this entanglement throughout the lattice.

Our earlier QLA simulations for electromagnetic wave propagation in dielectric media [1-4] were restricted to an initial one dimensional (1D) Gaussian pulse scattering from localized 2D dielectric structures and yielded very interesting transient interference effects. Because of imposed periodicity boundary conditions in QLA, our previous simulations were restricted to normal incidence when considering simple 1D initial pulses scattering rom infinite dielectric interfaces. Recently [14], by considering 2D spatially bounded initial profiles we were able to consider phenomenon like total internal reflection under initial value conditions. Unlike steady state simulation, QLA uncovered the Goos-Hanchen shift [15] in the reflected pulse - a spatial shift in the pulse along the dielectric interface. We saw how the transient fields in the second medium helped rotate the initial Poynting unit vector through twice the incident angle and form the reflected outgoing Poynting vector in total internal reflection.

In this paper, we will investigate the complementary regimes where the incident angle is less than the critical angle and look at transient effects as the pulse width interacts with the dielectric interface between two dielectric media. In Section 2 we briefly review how the Dyson map [5], [16] can generate a unitary representation for the inhomogeneous Maxwell equations. A unitary representation will permit direct encoding of the algorithm onto a quantum computer, as well as its norm will immediately yield an algorithm that exactly conserves the total electromagnetic energy throughout the simulation. In Section 3 the unitary collision and streaming operators that form the backbone of our QLA for Maxwell equations are presented. The derivatives of the dielectric functions of the medium are currently represented by a non-unitary but sparse matrix. For quantum computing, this non-unitary matrix can be written as a linear combination of unitary matrices (LCUs) [17], allowing for quantum encoding. In Section 4 we present some QLA simulations for oblique scattering when the refractive indices n1 < n2 and n1 > n2 as well as three different initial electromagnetic pulses: a burst pulse, also considered in [14] for total internal reflection studies, a thin elongated pulse (whose length is five times that of the burst pulse but width a fifth of that of the pulse burst) , and a pulse with intermediate geometric properties which we call a finite pulse. The magnetic field of these initial conditions are shown in 

\section{Maxwell Equations in Dyson Variables}
Consider the propagation of an electromagnetic pulse in a non-magnetic inhomogeneous dielectric medium with 
dielectric tensor $\mathcal{\epsilon}$.
The Maxwell equations for this system (in standard notation) are
\begin{equation}\label{Maxwell}
\begin{aligned}
\partial \mathbf{B}/ \partial t=-\nabla \times \mathbf{E} \\
\partial \mathbf{D} / \partial t=\nabla \times \mathbf{H} 
\end{aligned}
\end{equation}
with constitutive equations  $\mathbf{D} =  \mathcal{\epsilon}\cdot \mathbf{E}$ and  $\mathbf{B}=\mu_{0} \mathbf{H}$.
($\mathbf{E}$ is the electric field, $\mathbf{H}$ is the magnetic field.  $\mathbf{D}$ is the electric displacement and
$\mathbf{B}$ the magnetic induction.)
$\nabla \cdot \mathbf{D} = 0 = \nabla \cdot \mathbf{B}$ can be treated as initial conditions:  if they are satisified
initially then (in the continuum limit) they are satisfied for all time.

It is convenient to treat  $\mathbf{u}=(\mathbf{E}, \mathbf{H})^T$ as the basic field and 
$\mathbf{d} = (\mathbf{D}, \mathbf{B})^T$ as the derived field.  ${T}$ is just the transpose.
Thus, for a lossless medium in a given coordinate system aligned with its principal axes
\begin{equation} \label{W-def}
\mathbf{d} = \mathbf{W} \, \mathbf{u} \quad \text{with} \quad \mathbf{W}=\left[\begin{array}{cc}
\epsilon_i \mathbf{I}_{3 \times 3} & 0_{3 \times 3} \\
0_{3 \times 3}& \mu_{0} \mathbf{I}_{3 \times 3}
\end{array}\right] 
\end{equation}
$\mathbf{I}_{3 \times 3} $ is the 3D identity matrix.  For lossless media,
the block diagonal  $\mathbf{W}$ is Hermitian and invertible.  
The Maxwell equations, Eq. \eqref{Maxwell}, can then be written in the field $\mathbf{u}$
\begin{equation} \label{eqA1}
i \frac{\partial \mathbf{u}}{\partial t}=\mathbf{W}^{-\mathbf{1}} \mathbf{M} \mathbf{u} \quad \text{with} \quad
\mathbf{M}=\left[\begin{array}{cc}
0_{3 \times 3} & i \nabla \times \\
-i \nabla \times & 0_{3 \times 3}
\end{array}\right].
\end{equation}
Under standard boundary conditions, the curl-matrix operator $\mathbf{M}$ is Hermitian.  
By defining the operator $\mathbf{G}$
\begin{equation} \label{G-oph}
\mathbf{G} = \mathbf{W}^{-1} \mathbf{M} \quad , \quad \text{one has} \quad
i \frac{\partial \mathbf{u}}{\partial t}=\mathbf{G}\mathbf{u} .
\end{equation}
For a homogeneous dielectic  the Hermitian operators $\mathbf{W}^{-1}$ and $\mathbf{M}$ commute,
so that $\mathbf{G}^{\dagger} = \mathbf{M} \cdot \mathbf{W}^{-1} =  \mathbf{W}^{-1} \cdot \mathbf{M} = \mathbf{G}$ 
 : i.e., $\mathbf{G}$ is Hermitian.
The time evolution of $\mathbf{u}$, Eq. \eqref{G-oph}, is thus unitary.  However, for inhomogeneous dielectric media
$\mathbf{W}^{-1}$ and $\mathbf{M}$  no longer commute, $\mathbf{G} $ is not Hermitian,
and Eq. \eqref{G-oph} gives a non-unitary evolution for $\mathbf{u}$.

However,  a unitary representation for the inhomogeneous Maxwell equations can be found
provided this matrix $\mathbf{G}$, Eq. \eqref{G-oph}, is pseudo-Hermitian [5, 16] : i.e.,
provided there exists a Hermitian operator $\mathbf{K}$ such that
\begin{equation} \label{pseudo}
\mathbf{G}^{\dagger} = \mathbf{K} \cdot \mathbf{G} \cdot \mathbf{K}^{-1} \quad .
\end{equation}
But for the operator $\mathbf{G}$, Eq. \eqref{G-oph}, we can readily determined the required Hermitian $\mathbf{K}$
since
\begin{equation} \label{G-psuedo}
\mathbf{G}^{\dagger} = \mathbf{M} \cdot \mathbf{W}^{-1} = (\mathbf{W}.\mathbf{W}^{-1}) .\mathbf{M} \cdot \mathbf{W}^{-1} = \mathbf{W}.(\mathbf{W}^{-1} \mathbf{M}).\mathbf{W}^{-1} = \mathbf{W}. \mathbf{G}.\mathbf{W}^{-1}  \quad .
\end{equation}
i.e., $\mathbf{G} = \mathbf{W}^{-1} \mathbf{M} $ is pseudo-Hermitian under the constitutive Hermitian operator 
$\mathbf{W}$, Eq. \eqref{W-def}.
Moreover it can be readily shown
that the operator $\mathbf{G}$ is actually Hermitian under the inner product defined with the weighting function 
$\mathbf{W}$.  Indeed, Since $\mathbf{M}$ and $\mathbf{W}$ are Hermitian,
\begin{equation}
\begin{aligned}
\bra{\mathbf{u}_1} \ket{\mathbf{G.u_2}}_\mathbf{W} = \bra{\mathbf{u}_1} \ket{\mathbf{W.G.u_2}}
 = \bra{\mathbf{u}_1} \ket{\mathbf{M.u_2}}  = \bra{\mathbf{M u_1 }} \ket{\mathbf{u_2}}  \\
 =\bra{\mathbf{W.G. u_1 }} \ket{\mathbf{u_2}}  = \bra{\mathbf{G. u_1} }\ket{\mathbf{u_2}}_\mathbf{W}  .
\end{aligned}
\end{equation}

Therefore a Dyson map [5], [10]-[12] $\mathbf{u} \rightarrow \mathcal{U}$
can be inferred that preserves the inner product structure between the initial and transformed representations:
\begin{equation}
\bra{ \mathbf{u}_1} \ket {\mathbf{u}_2}_\mathbf{W} = \bra{\mathcal{U}_1} \ket{\mathcal{U}_2} \quad , \quad
\text{for some }  \mathcal{U} .
\end{equation}
A simple choice comes from 
\begin{equation}
\bra{ \mathbf{u}_1} \ket {\mathbf{u}_2}_\mathbf{W}  = \bra{ \mathbf{u}_1} \ket {\mathbf{W} \,\mathbf{u}_2} =
\bra{ \mathbf{W}^{1/2} \mathbf{u}_1 } \ket {\mathbf{W}^{1/2} \,\mathbf{u}_2} 
\equiv \bra{\mathcal{U}_1} \ket{\mathcal{U}_2} ,
\end{equation}
where $\mathcal{U}_i \equiv \mathbf{W}^{1/2} \mathbf{u_i}, \, i=1,2$.
Thus an appropriate Dyson map for Maxwell equations is
\begin{equation} \label{eqA3}
\mathcal{U} = \mathbf{W}^{1/2} \mathbf{u} ,
\end{equation}
where $\mathbf{u} = \mathbf{(E, H)}^T$.  Thus the inhomogeneous Maxwell evolution equations, under the Dyson map, 
are transformed from the non-unitary evolution for 
$\mathbf{u} = \mathbf{(E, H)}^T$ to the unitary evolution for $\mathcal{U}$
\begin{equation} \label{eqA4}
i \frac{\partial \mathbf{u}}{\partial t}=\mathbf{W}^{-\mathbf{1}} \mathbf{M} \mathbf{u} \qquad  \rightarrow  \qquad
i \frac{\partial \mathcal{U}}{\partial t}=\mathbf{W}^{-1 / 2} \mathbf{M} \mathbf{W}^{-1 / 2} \mathcal{U}
\end{equation}
\noindent since $\mathbf{W}^{-\mathbf{1} / \mathbf{2}} \mathbf{M} \mathbf{W}^{-\mathbf{1} / \mathbf{2}}$ is Hermitian.
For non-magnetic materials, Eq. \eqref{eqA3} is 
\begin{equation} \label{eqA5}
\mathcal{U}=\left(\epsilon_0^{1/2} n_x E_x, \epsilon_0^{1/2}  n_y E_y,  \epsilon_0^{1/2} n_z E_z, \mu_0^{1/2}  \mathbf{H}\right)^{T}    .
\end{equation}
where $(n_x , n_y, n_z)$ is the vector (diagonal) refractive index, with $\epsilon_i =\epsilon_0  n_i^2, i=x,y,z$ .
The unitary evolution Eq. \eqref{eqA4} for 2D $x-y$ spatially dependent fields and dielectrics is
\begin{equation} \label{eqA6}
\begin{aligned}
\frac{\partial q_0}{\partial t} = \frac{1}{n_x} \frac{\partial q_5}{\partial y} , \qquad
\frac{\partial q_1}{\partial t} = - \frac{1}{n_y} \frac{\partial q_5}{\partial x} , \qquad
\frac{\partial q_2}{\partial t} =  \frac{1}{n_z} \left[ \frac{\partial q_4}{\partial x} -\frac{\partial q_3}{\partial y} \right] \\
\frac{\partial q_3}{\partial t} = - \frac{\partial (q_2/n_z)}{\partial y} , \qquad
\frac{\partial q_4}{\partial t} = \frac{\partial (q_2/n_z)}{\partial x} , \qquad
\frac{\partial q_5}{\partial t} = - \frac{\partial (q_1/n_y)}{\partial x}  + \frac{\partial (q_0/n_x)}{\partial y}   ,
\end{aligned}
\end{equation}
with $\mathcal{U} \equiv (q_0, q_1, q_2, q_3, q_4, q_5)^T$ being the qubit amplitudes.

\subsection*{Conservation of Instantaneous Total Electromagnetic Energy }
While unitary evolution is essential for direct quantum encoding, its importance in  classical computations
is the enforcement of the norm of the Dyson variable $\mathcal{U}$ to be a constant of the motion.  This constant is immediately identified to be the
the total electromagnetic energy  $\mathcal{E}(t)$ for our Maxwell equations:
\begin{equation} \label{eqA13}
\mathcal{E}(t) =  \int_0^L \int_0^L dx dy \left[ \epsilon_0( n_x^2 E_x^2 + n_y^2 E_y^2 + n_z^2 E_z^2 )+  \mu_0 \mathbf{H}^2 \right] 
= ||\, \mathcal{U}||^2 = const.
\end{equation}  
where the total system is in a square box of length $L$.

\section{QLA for Eq. \eqref{eqA6}}
The difficulty with determining a QLA for a particular physics problem is in finding the specific unitary collision operators and streaming operators required as well as interleaving them among themselves on a discrete lattice so that
in the discrete $\rightarrow$ continuum limit one recovers the required continuum partial differential equations of interest
to second order in the lattice spacings. As the detailed QLA matrices and sequences to recover Eq. \eqref{eqA6}
have been presented in detail elsewhere [1-4, 14], we shall here discuss some of the issues that occur.

The simplest QLA scheme for a set of first order partial differential equations, as is Eq. \eqref{eqA6}, is to work 
independently with the orthogonal Cartesian directions.  We first consider the evolution of the qubit amplitudes that 
couple the spatial derivatives on the other amplitudes.  For the $\partial / \partial y$-terms there is the coupling
of $q_0 \leftrightarrow q_5$ and $q_2 \leftrightarrow q_3$.  These sub-couplings are $2 \times 2$ unitary matrices, so 
that the full $6 \times 6$ unitary collision matrix required has the form
\begin{equation} \label{eqA7}
C_Y=\left[\begin{array}{cccccc}
cos \,\theta_0& 0 & 0 & 0& 0& - sin\,\theta_0 \\
0 & 1 & 0 & 0 & 0 &0\\
0 & 0 & cos\,  \theta_2& - sin \,\theta_2 & 0 & 0 \\
0 & 0 & sin \,\theta_2 & cos\,  \theta_2 &  0 & 0  \\
0 & 0 & 0 & 0 & 1 & 0 \\
sin\,\theta_0 & 0& 0 & 0 & 0 & cos \,\theta_0
\end{array}\right]
\end{equation}
The collision angle $\theta_0$ is a function of refractive index $n_x$ while $\theta_2$ is a function of $n_z$.
The unitary streaming operator to recover the spatial derivative $\partial / \partial y$ will stream the amplitudes
$q_0 \leftrightarrow q_3$ and $q_2 \leftrightarrow q_5$, while all the other amplitudes remain on their lattice sites
unstreamed.

Finally, to recover the spatial derivatives on the refractive indices we must introduce potential interaction terms.
To recover the y-spatial inhomogeneity we have to couple amplitudes $q_3 - q_2$ and $q_5 - q_0$.  
An appropriate y-potential operator is
\begin{equation} \label{eqA9}
V_Y=\left[\begin{array}{cccccc}
1 & 0 & 0 & 0& 0& 0 \\
0 & 1 & 0 & 0 & 0 & 0\\
0 & 0 & 1 & 0 &0& 0 \\
0 & 0 & - sin\,\beta_2  & cos\, \beta_2 &  0 & 0  \\
0 & 0 & 0& 0 & 1 & 0 \\
sin\, \beta_0& 0 & 0 & 0 & 0 & cos \,\beta_0
\end{array}\right]
\end{equation}
with suitably chosen angles $\beta_0$ and $\beta_2$.  Clearly $V_Y$ is a non-unitary matrix.  
Similarly when determining the potential operator for the x-inhomogeneity in the refractive index, this
operator is likewise non-unitary. 

When running QLA on a classical computer, the non-unitarity of $V_Y$ creates no coding problems, but since
the classical QLA  is not fully unitary, we are not guaranteed conservation of energy, Eq. \eqref{eqA13} for
all time steps.  To recover a second order numerical scheme for the solution of Eq. \eqref{eqA6}, we need
an interleaved product of 32 unitary operators along with the 2 non-unitary potential operators $V_X, V_Y$:
i.e., an almost unitary representation of Eq. \eqref{eqA13}.
  In our classical QLA simulations, we thus need to monitor the total energy $\mathcal{E}_{QLA}(t)$ and it is
found in all simulations to remain a constant to the $7th$ significant figure throughout the run.  $\nabla .
\mathbf{H} = 0 $ to machine accuracy, while $O(\nabla . \mathbf{E}/max(\mathbf{E})) < 10^{-17}$.

To run this QLA on a quantum computer one must handle non-unitary matrices like
 $V_Y$.  Non-unitary matrices are
  a problem well-studied in the field of quantum information sciences - particularly when
one is dealing with open quantum systems, i.e., when the system one is interested in is interacting and 
exchanging energy with the environment.   One method is to rewrite the non-unitary matrix as a linear 
superposition of  unitary matrices.  (for $V_Y$ it will be 4 unitary matrices), and then follow the much
published efforts on putting a linear combination of unitaries (LCU) onto a quantum computer [17]-[19].  
Another approach is to determine analytically the singular value decomposition of the non-unitary matrix 
$V_Y$
into a product of 3 matrices :  $V_Y=\mathcal{A}.\mathcal{D}.\mathcal{B}$, where $\mathcal{A}$ and 
$\mathcal{B}$ are unitary, and $\mathcal{D}$ is non-unitary but diagonal.   One can immediately
renormalize the diagonal elements by the maximum eigenvalue of $\mathcal{D}$ so that all the eigenvalues
are bounded by one.  This renormalized diagonal matrix can be immediately decomposed into the sum 
of 2 unitary matrices and one is back into the well-studied subject of implementing LCUs onto a quantum computer.

\section{QLA Simulations for Oblique Angle Scattering by a Plane Dielectric Interface}
\qquad It is of interest to consider oblique angle scattering of three different pulse shapes for an incident angle less than
the critical angle of total internal reflection (when medium index $n_1 > n_2$). In [14] we have considered the
interesting case when the incident angle is such that there is total internal reflection.  In this case, the transient 
fields in the vacuum region $n_2$ played a pivotal role in rotating the incident Poynting flux vector to the reflected
vector along with a Goos-Hanchen [15] spatial shift along the dielectric interface.

The plane of incidence is defined by the standard Cartesian $(x,y)$ coordinates with the two dielectric regions ${0 \le x < L, 0 \le y < L/2}$
and ${0 \le x < L, L/2 \le y < L}$.  For $p$-polarization [20], the incident electric field, $\mathbf{E}$, is in the $x-y$ plane and the corresponding
magnetic field, $\mathbf{H}$, is in the $-z$-direction.
For the initial 2D pulse we use its natural coordinate system - a rotated Cartesian system $(\zeta, \chi)$ where the direction of incident
propagation is $\mathbf{\hat{\zeta}}$, the magnetic field is in the $-\hat{z}$ direction, and the  electric field in the $\hat{\chi}$
direction.  An appropriate incident magnetic field amplitude is 
\begin{equation} \label{eq1}
H_z(\zeta,\chi) = - exp \left[- \left( \frac{\zeta - \zeta_0}{\zeta_w} \right)^2  -  \left( \frac{\chi - \chi_0}{\chi_w} \right)^2  \right]  cos \left(\frac{2 \pi \zeta}{\gamma_w} \right) .
\end{equation}
Here $\zeta_w$ defines the extent of the packet in the $\zeta$-direction, and similarly $\chi_w$ in the $\chi$-direction. 
The two coordinate systems are related by the rotation matrix
 \begin{equation}\label{coordss}
\begin{bmatrix}
x \\
y
\end{bmatrix}=\begin{bmatrix}
cos \,\theta & sin \, \theta \\
-sin \, \theta & cos \,\theta  
\end{bmatrix}\begin{bmatrix}
\zeta \\
\chi
\end{bmatrix}
\end{equation}
where $\theta$ is the angle of incidence in the $(x,y)$-plane.

 The corresponding electric field amplitude $E_{\chi}(\zeta,\chi)$ is related to the magnetic field amplitude $H_z$
by the impedance of the medium, $Z=E_{\chi}/H_z$.

We consider the following initial conditions, with the same angle of incidence $\theta = 25^o < \theta_c = 30^o$ 
where $\theta_c$ is the critical angle for onset of total internal reflection (when $n_1=2, n_2=1$):  

\begin{table}[ht]
\centering
\begin{tabular}{|c|c|c||c|c|c|c|} \hline
Figure         & $ n_1$  &  $n_2$  &  pulse  & $\zeta_w$ & $\chi_w$  & $\gamma_w$ \\ \hline \hline
Fig. 1a & 1 & 2  & burst & 20  &  100 &  20  \\  \hline
Fig. 1b & 1 & 2  & thin, long & 100  &  20  &  20    \\  \hline  \hline
Fig. 2a & 2 & 1  &  burst & 20  &  100 &  20    \\   \hline
Fig. 2b & 2 & 1   & thin, long & 100  &  20  &  20   \\  \hline
Fig. 3 & 2 & 1  & finite   & 50  &  50  &  20  \\  \hline  \hline
\end{tabular}
\caption{\it{Parameters for the QLA simulations on a $1024 \times 1024$ grid.}
}
\end{table}

The time evolution of the oblique scattering from the incident region with $n_1=1$ to the transmitted region with
$n_2=2 > n_1$ is shown in Fig. 1.  The initial burst profile, Fig. 1(a.0), interacts with the interface at t = 20k iterations, 
Fig. 1(a.20) with a dominant transmission and weak reflected pulses at t = 32k, Fig. 1(a.32).  The
transmitted pulse's wavelength is now half the incident wavelength since region $n_2=2$, Figs. 1(a.44) and
1(a.56).  The reflected burst retains its initial wave form.  

For an initial thin but elongated  pulse, Fig. 1(b.0),
following the somewhat complex interference effects [Figs 1(b.20) and 1(b.32),  
the resulting reflected pulse resembles wave fronts emitted from a point-like source at the dielectric interface,
 Figs. 1(b,44) and (b.56).  The transmitted pulse is compressed (because of the lower
 refractive index $n_2$) but with somewhat more width.

 \begin{figure}[!h!p!b!t] \ 
\begin{center}
\includegraphics[width=3.2in,height=2.0in]{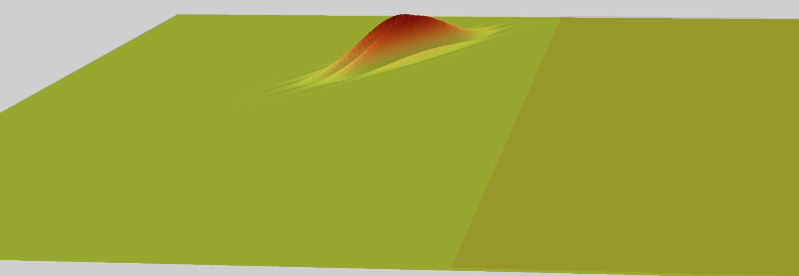}
\includegraphics[width=3.2in,height=2.0in]{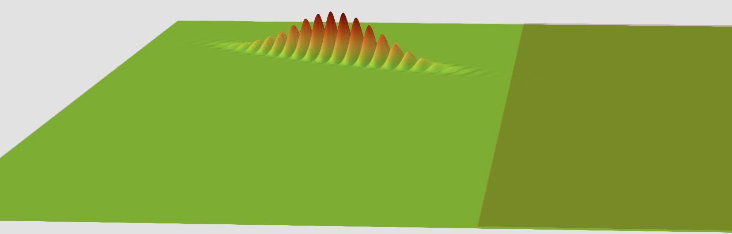}
 $ t = 0 : \qquad$ (a.0) burst \qquad    \qquad  \qquad \qquad \qquad (b.0)  thin, long pulse\\
\includegraphics[width=3.2in,height=2.0in]{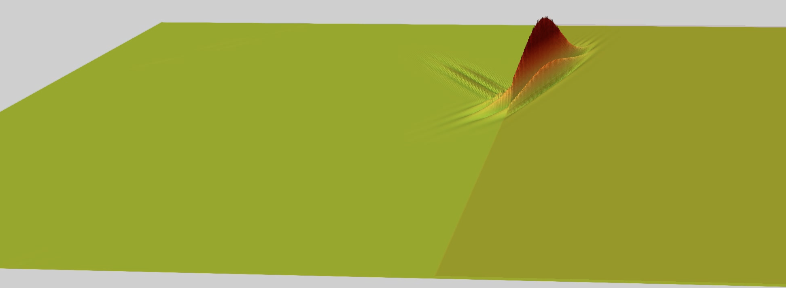}
\includegraphics[width=3.2in,height=2.0in]{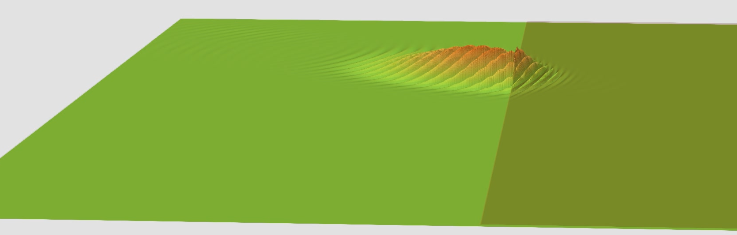}
\qquad $ t = 20k :\qquad  $ (a.20) burst   \qquad  \qquad   \qquad  \qquad  \qquad (b.20)  thin, long pulse\\\
\includegraphics[width=3.2in,height=2.0in]{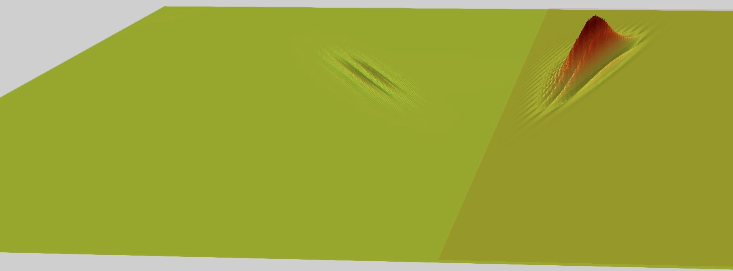}
\includegraphics[width=3.2in,height=2.0in]{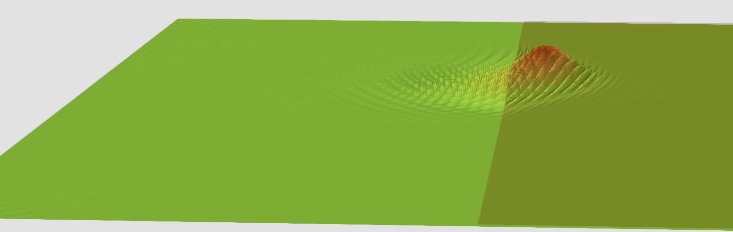}
\qquad   $ t = 32k : \qquad  $ (a.32) burst  \qquad \qquad   \qquad  \qquad \qquad (b.32)  thin, long pulse\ \\
\end{center}
\end{figure}

 \begin{figure}[!h!p!b!t] \ 
\begin{center}
\includegraphics[width=3.2in,height=2.0in]{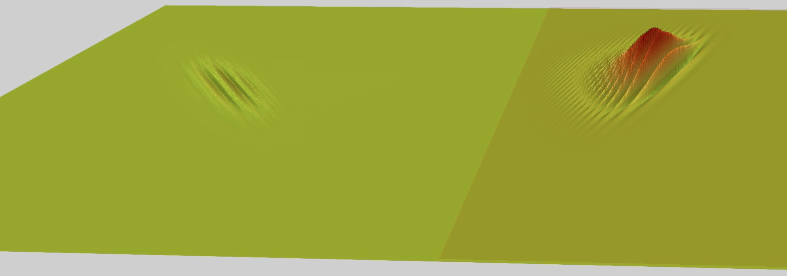}
\includegraphics[width=3.2in,height=2.0in]{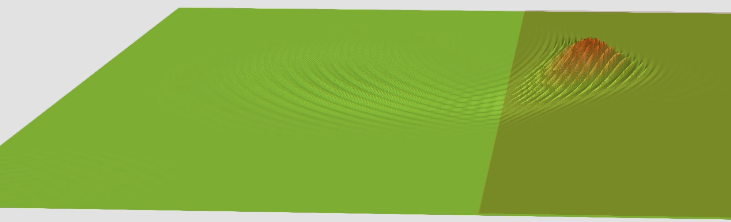}
\end{center}
$ t = 44k $ \qquad (a.44)  burst \qquad \qquad   \qquad  \qquad  \qquad (b.44)  thin, long pulse\\
\includegraphics[width=3.2in,height=2.0in]{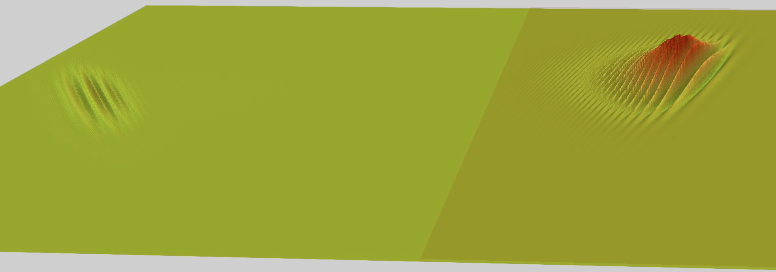}
\includegraphics[width=3.2in,height=2.0in]{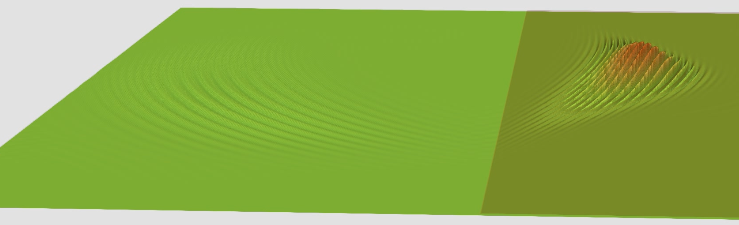}
$ t = 56k $ \qquad (a.56) burst \qquad   \qquad  \qquad  \qquad \qquad (b.56)  thin, long pulse \\
\caption{\it{Evolution of the magnetic field $H_z(x,y) > 0$  for incident angle $\theta = 25^o < \theta_c$ for 
two different pulse shapes.
$ n_1=1 \text{(left side)} \rightarrow n_2=2 \text{(right side)}$.  Notation:  Fig 1(b.56) in the text refers to
Fig. 1b (thin long pulse) at time t = 56k.}
}
\end{figure}

Somewhat more interesting are the transmitted and reflected $H_z >0)$ profiles when the scattering occurs
from the high to low refractive indices, Fig. 2 - even for scattering angles below the critical angle for total internal
reflection.  Consider first the burst $H_z > 0$ profile, Fig. 2(a.0).  As the pulse interacts with the interface, we now see 
the birth of the transmitted pulse [Fig.2(b.20)] and by time t = 32k one sees the reflected pulse as it moves back 
into the $n_1=2$ dielectric and its asymptotic profile in Fig. 2(a.60) and Fig. 2(a.76k).  The transmitted pulse, as it
propagates deeper into the $n_2=1$ dielectric, has double the wavelength but its Gaussian envelope elongates 
perpendicular to the direction of propagation, Figs. 2(a.32), 2(a.60) and 2(a.76).  
For an initial thin elongated pulse, Fig. 2(b.0), there is the slow formation of the reflected pulse with similar
characteristics to the initial pulse [Figs. 2(b.20), 2(b.32), 2(b.60) and 2(b.76)].
The transmitted pulse, however, takes on a wavefront pattern reminiscent of that wavefronts emitted by a point-like
source at the interface.

 \begin{figure}[!h!p!b!t] \ 
\begin{center}
\includegraphics[width=3.1in,height=2.0in]{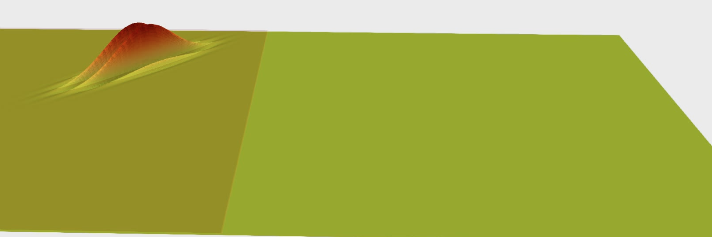}
\includegraphics[width=3.1in,height=2.0in]{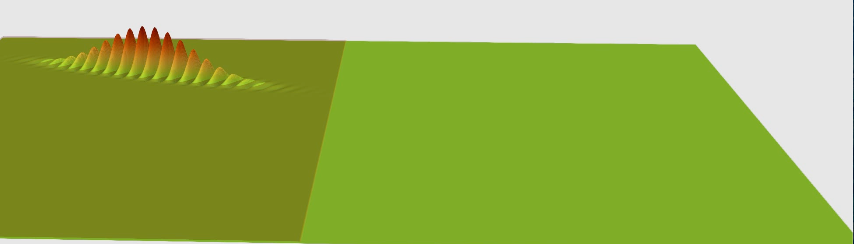}
$ t = 0 $ (a.0) burst \qquad   \qquad  \qquad  (b.0)  thin, long pulse \\
\includegraphics[width=3.1in,height=2.0in]{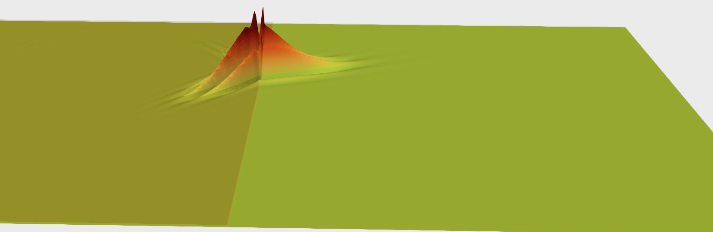}
\includegraphics[width=3.1in,height=2.0in]{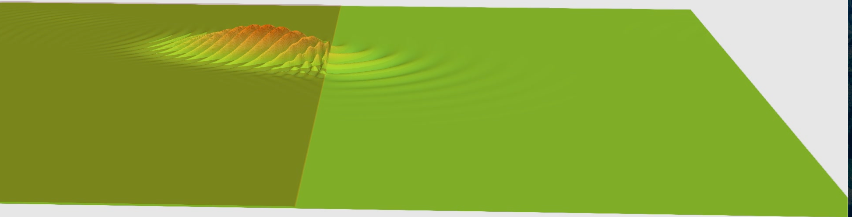}
$ t = 20k :$ (a.20) burst \qquad   \qquad  \qquad  (b.20) thin, long pulse \\
\includegraphics[width=3.1in,height=2.0in]{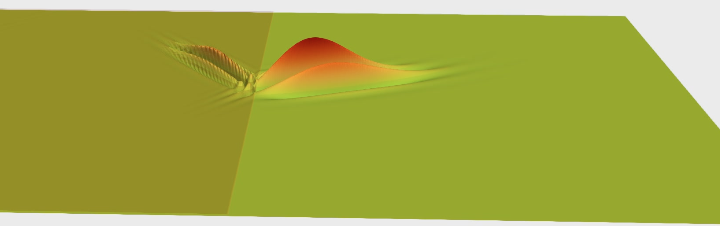}
\includegraphics[width=3.1in,height=2.0in]{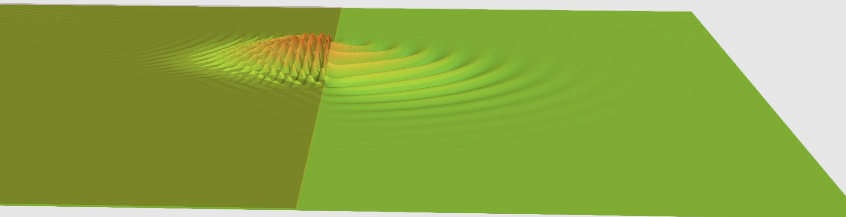}
$ t = 32k : $ (a.32) burst \qquad   \qquad  \qquad   (b.32)  thin, long pulse \\
\end{center}
\end{figure}

 \begin{figure}[!h!p!b!t] \ 
\begin{center}
\includegraphics[width=3.2in,height=2.0in]{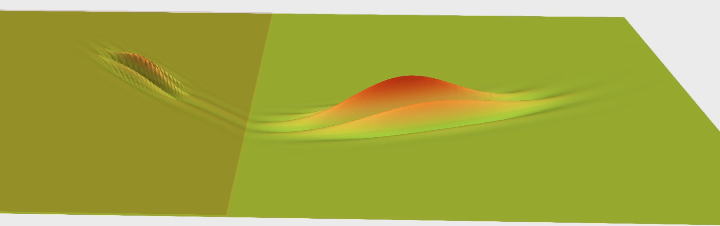}
\includegraphics[width=3.2in,height=2.0in]{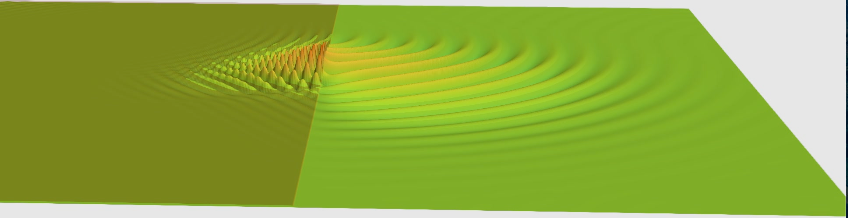}
$ t = 60k : $ (a.60) burst \qquad   \qquad  \qquad  (b.60)  thin, long pulse \\
\includegraphics[width=3.2in,height=2.0in]{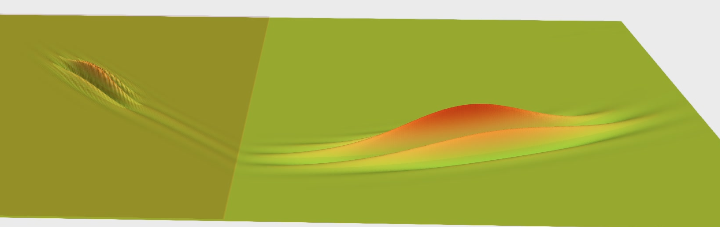}
\includegraphics[width=3.2in,height=2.0in]{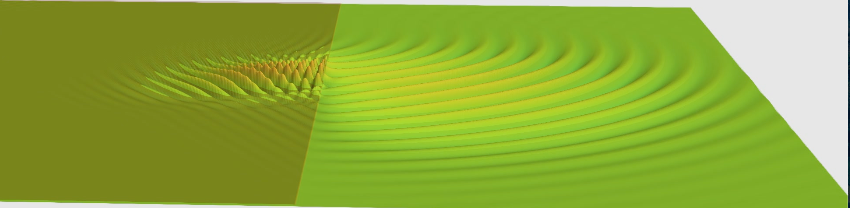}
$ t = 76k : $ (a.76) burst\qquad   \qquad  \qquad  (b.76)  thin, long pulse \\
\caption{\it{Evolution of the magnetic field $H_z(x,y) > 0$  for incident angle $\theta = 25^o < \theta_c=30^o$ for 
two different pulse shapes.
$ n_1=2 \text{(left side)} \rightarrow n_2=1 \text{(right side)}$.  Notation:  Fig 2(a.76) in the text refers to
Fig. 2a (burst) at time t = 76k.}
}
\end{center}
\end{figure}

These ideas are further reinforced by considering the transient behavior of a finite pulse, FIg. 3 (a) , in which
the pulse is longer than the burst pulse, Fig.  1a, but shorter than the
thin long pulse, Fig. 1b.  The finite pulse 
has more oscillations than the burst pulse, Fig. 1a, but less than the thin long pulse, Fig. 1b.  The transient fields have 
similar intermediate behavior:  the reflected pulse is weaker and slightly more diffuse while the transmitted pulse
exhibits both a significant Gaussian feature that is quite diffuse and the Huygen wavefronts as for a localized
source at the interface. 

 \begin{figure}[!h!p!b!t] \ 
\begin{center}
\includegraphics[width=3.2in,height=2.0in]{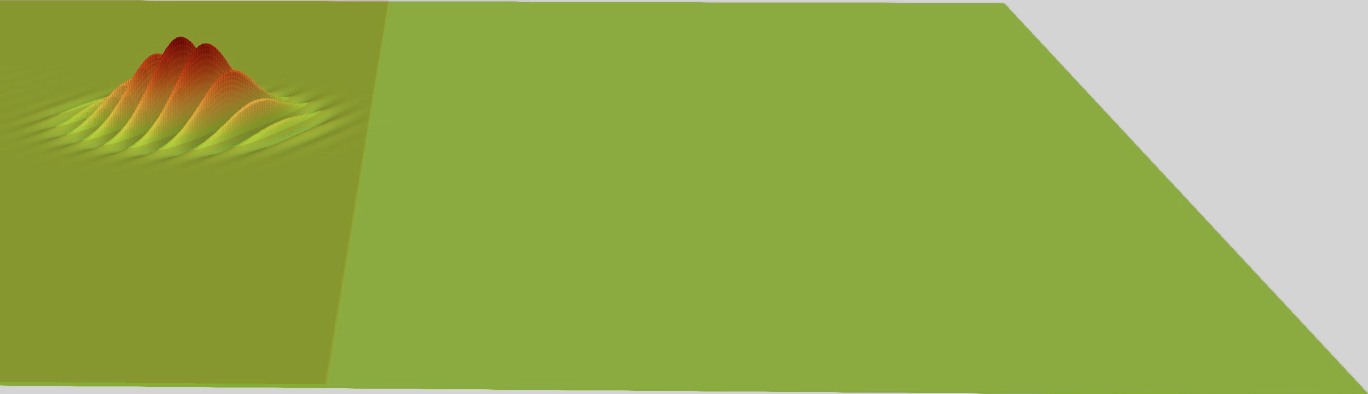}
\includegraphics[width=3.2in,height=2.0in]{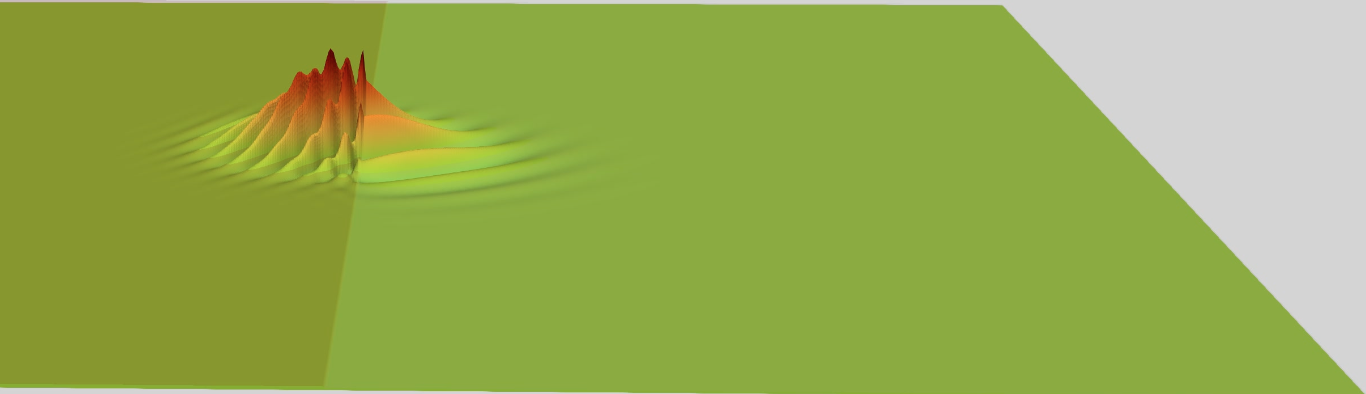}
 finite pulse    (a)  t = 0k :    \qquad   \qquad  \qquad  (b)  t =   15k \\
\includegraphics[width=3.2in,height=2.0in]{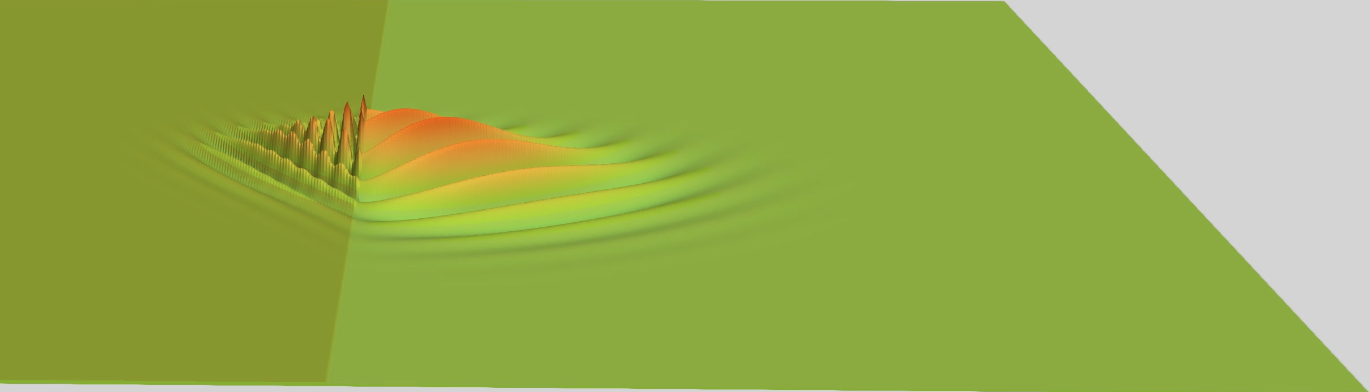}
\includegraphics[width=3.2in,height=2.0in]{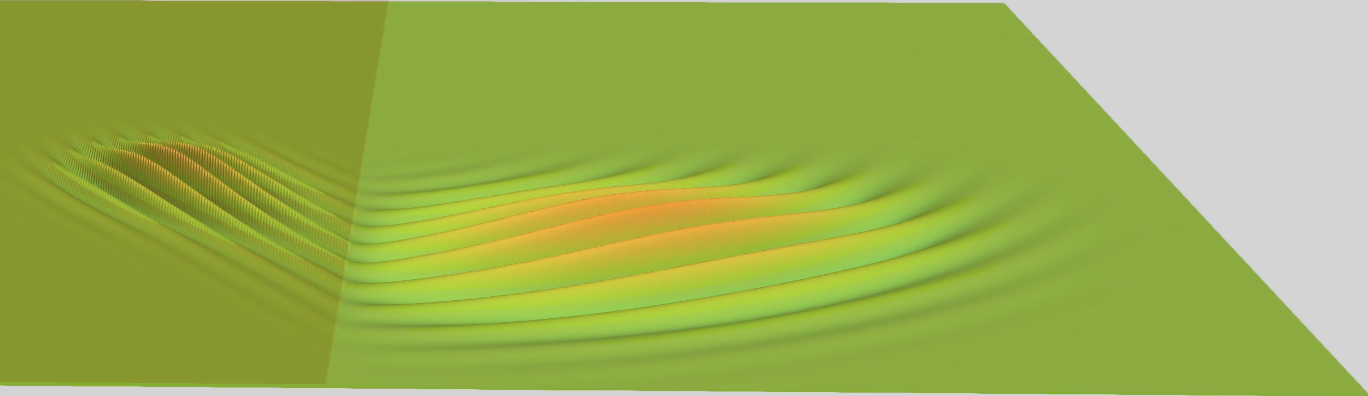}
 finite pulse    (c)  t = 30k :    \qquad   \qquad  \qquad  (d)  t =   45k \\
\caption{\it{Evolution of the magnetic field $H_z(x,y) > 0$  for incident angle $\theta = 25^o < \theta_c=30^o$ for 
a finite pulse.
$ n_1=2 \text{(left side)} \rightarrow n_2=1 \text{(right side)}$. }
}
\end{center}
\end{figure}

\section{Summary and Conclusions}

\qquad Here we have examined the transient fields that occur when an obliquely incident electromagnetic pulse
interacts with an infinite planar dielectric interface, using our second order accurate QLA Maxwell code.  
In particular we have looked at the effect of the width and length of these Gaussian-like pulses on the reflected and
transmitted pulses.  For the transmitted pulse one finds a combination of a Gaussian-like pulse superimposed on
Huygen point wavefronts.  The thinner the pulse, the  point-like Huygen wavefronts 
become more dominate in the transmitted pulse.

Because of the importance currently placed on quantum algorithms having minimal circuit depth, 
many quantum differential
equation solvers remain as first order schemes in the discreteness parameter.  For nonlinear problems - besides the
severe headache of how to handle time-evolving nonlinearities in  linear quantum mechanics without resorting to
a hybrid classical-quantum computer and its complexities - our look at the very simple Lorenz nonlinear equations
[21] indicates that accurate simulations would need to go to second order schemes.  QLA is a second order scheme -
and in the time evolution step one evaluates an interleaved sequence of 32 unitary collision-streaming operators
and just 2 non-unitary but very sparse matrices, 
(For a first order QLA scheme one would need only 16 collide-stream matrix products with the 2 non-unitary 
potential operators).
A fully unitary QLA representation would automatically enforce to
machine accuracy the total electromagnetic energy of the system - and a search for such an algorithm is
underway as this would be also most beneficial for QLA run on a classical computer.

\section*{Acknowledgments}
This research was partially supported by Department of Energy grants DE-SC0021647, DE-FG02- 91ER-54109, DE-SC0021651, DE-SC0021857, and DE-SC0021653. This work has been carried out within the framework of the EUROfusion Consortium, funded by the European Union via the Euratom Research and Training Programme (Grant Agreement No. 101052200 - EUROfusion). Views and opinions expressed, however, are those of the authors only and do not necessarily reflect those of the European Union or the European Commission. Neither the European Union nor the European Commission can be held responsible for them. E. K. was supported by the Basic Research Program, NTUA, PEVE. K.H is supported by the National Program for Controlled Thermonuclear Fusion, Hellenic Republic. This research used resources of the National Energy Research Scientific Computing Center (NERSC), a U.S. Department of Energy Office of Science User Facility located at Lawrence Berkeley National Laboratory, operated under Contract No. DE-AC02-05CH11231 using NERSC award FES-ERCAP0020430.

\section*{REFERENCES}

\qquad [1]  G. Vahala, L. Vahala, M. Soe, and A. K. Ram, "Unitary quantum lattice simulations for Maxwell equations in vacuum and in dielectric media," J. Plasma Phys. 86, 905860518 (2020).

[2]  G. Vahala, L. Vahala, M. Soe, and A. K. Ram, "One- and two dimensional quantum lattice algorithms for Maxwell equations in inhomogeneous scalar dielectric media I: theory," Radiat. Eff. Defects Solids 176, 49?63 (2021). 

[3]  G. Vahala, J. Hawthorne, L. Vahala, A. K. Ram, and M. Soe, "Quantum lattice representation for the curl equations of Maxwell equations," Radiat. Eff. Defects Solids 177, 85?94 (2022). 

[4]  G. Vahala, M. Soe, L. Vahala, A. K. Ram, E. Koukoutsis, and K. Hizanidis, "Qubit lattice algorithm simulations of Maxwell's equations for scattering from anisotropic dielectric objects," Comput. Fluids 266, 106039 (2023).

[5]  E. Koukoutsis, K. Hizanidis, A. K. Ram, and G. Vahala, "Dyson maps and unitary evolution for Maxwell equations in tensor dielectric media," Phys. Rev. A 107, 042215 (2023)

[6]  G. Vahala, M. Soe, E. Koukoutsis, K. Hizanidis, L. Vahala, and A. K. Ram, "Qubit lattice algorithms based on the Schrodinger-Dirac representation of Maxwell equations and their extensions," in 'Schrodinger Equation - Fundamentals Aspects and Potential Applications', edited by D. M. B. Tahir, D. M. Sagir, A. P. M. I. Khan, D. M. Rafique, and D. F. Bulnes (IntechOpen, Rijeka, 2023) Chap. 5.

[7]  J. Yepez, "A quantum lattice-gas model for computational fluid dynamics," Phys. Rev. E63, 046702 (2001).

[8]  J. Yepez, "Quantum lattice-gas model for Burgers equation", J. Stat. Phys. 107, 203-224 (2002)

[9]  J. Yepez, "Relativistic path integral as a lattice-based quantum algorithm", Quantum Infor. Proc. 4, 471-509 (2005)

[10]  M. A. Pravia, Z. Chen, J. Yepez and D. G. Cory, "Experimental Demonstration of Quantum Lattice Gas Computation", arXiv:0303183
(2003)

[11]  J. Yepez, "Quantum lattice gas model of Dirac particles in 1+1 dimensions", arXiv:1307.3595 (2013)

[12]  E. Koukoutsis, K. Hizanidis, G. Vahala, M. Soe, L. Vahala, and A. K. Ram, "Quantum computing perspective
for electromagnetic wave propagation in cold magnetized plasmas", Phys. Plasmas 30, 122108 (2023).

[13] E. Koukoutsis, K. Hizanidis, A. K. Ram, and G. Vahala, "Quantum simulation of dissipation for Maxwell equations
in dispersive media", Future Gener. Comput. Syst. 159, 221 (2024).

[14]  M. Soe, G. Vahala, L. Vahala, E. Koukoutsis, A. K. Ram and K. Hizanidis, "Qubit lattice algorithm simulations of the scattering of a 
bounded two dimensional electromagnetic pulse from an infinite planar dielectric interface, arXiv:2512.02856 (2025)

[15]  R. H. Renard, "Total reflection: A new evaluation of the Goss-Hanchen shift", J, Optical Soc. America 54, 1190-1197 (1964);
F. Goos and H. Hanchen, Ann. Phys. (Leipzig) 436, 333 (1947).

[16]  A. Mostafazadeh, "Pseudo-Hermitian Representation of Quantum Mechanics", Int. J. Geom. Meth. Mod. Phys. 07, 1191 (2010)

[17]   A. M. Childs and N. Wiebe, "Hamiltonian simulation using linear combinations of unitary operations", Quantum Info. Comput. 12, 901?924 (2012).

[18]  M. Znojil, Quantum mechanics using two auxiliary inner products, Phys. Lett. A 421, 127792 (2022).

[19] A. Fring and M. H. Y. Moussa, Unitary quantum evolution for time-dependent quasi-Hermitian systems with nonobservable Hamiltonians, Phys. Rev. A 93, 042114 (2016).

[20]  A. Zangwill, "Modern Electrodynamics", Cambridge Univ Press, 2012.

[21]  E. Koukoutsis, G. Vahala, M. Soe, K. Hizanidis, L. Vahala and A. K. Ram, "Time-marching quantum 
algorithm for simulation of nonlinear Lorenz dynamics," Entropy 27, 871 (2025)

\end{document}